# Reply on the comment on the paper "Superconducting transition in Nb nanowires fabricated using focused ion beam"

G C Tettamanzi[1,2]*, A Potenza[3], S. Rubanov[4], C H Marrows[4] and S Prawer[1]

[1] School of Physics, University of Melbourne, Victoria 3010, Australia

[2] Kavli Institute of Nanoscience, TU Delft, Lorentzweg 1, 2628 CJ DELFT, The Netherlands

[3] School of Physics and Astronomy, University of Leeds, Leeds LS2 9JT, United Kingdom

[4] Bio21 Institute, University of Melbourne, Australia

*E-mail: G.C.Tettamanzi@tudelft.nl

**Abstract.**
In this communication we present our response to the recent comment of A. Engel regarding our paper on FIB-fabricated Nb nanowires (see Vol. **20** (2009) Pag. 465302). After further analysis and additional experimental evidence, we conclude that our interpretation of the experimental results in light of QPS theory is still valid when compared with the alternative proximity-based model as proposed by A. Engel.

Communication

In the comment of A. Engel to our recent work on FIB-fabricated Nb Nanowires [1], an alternative model, which does not account for quantum phase slip (QPS) was proposed [2] to interpret our data. This model is simply based on the proximity effect in an NbGa(sheath)/Nb(core) nanowire [2]. In particular, figure 1 in said comment [2] illustrates the dependence of the critical temperature $T_C$ with the thickness of the Nb core, $d_S$, as measured by HR-TEM, as well as the expected behaviour of $T_C$ versus $d_S$, calculated with the parameters for a Nb/Cu system. The inferred behaviour for our sample A (where we observe QPS and which comprises an extrapolated Nb core of 15 nm) according to the above model would be that of a superconducting transition between 3.5 -4 K and therefore not observed in the experimental range of our figure 2 in [1]. In this regard we like to remark that the depression of $T_C$ with $d_S$ in a proximity system is strongly dependent upon the conductivity of the non-superconducting material. Ideally a superconductor/insulator (SC/I) system would show no proximity effect at all and consequently no (or very weak) dependence of $T_C$ on $d_S$. On general grounds, Cu can be assumed as a much better conductor (i.e. having a lower resistivity) than Nb-Ga [4] especially for this Nb-Ga that is implanted with Ga atoms [4]. To be more clear, in the model used by Engel (see ref 4 in [2]), the resistivity of the Cu is estimated to be 1.3 μΩcm that is one order of magnitude lower than the resistivity of our Nb films (14.4 μΩcm, see ref [3]) and it is most likely true that the resistivity of the Nb-Ga regions of our nanowires will even higher than this (see ref [4]). Consequently the values of $T_C$ in the aforementioned figure 1 for our Nb/Nb-Ga system can be expected to be considerably higher than the ones calculated by Engel and therefore within the experimental range of our figure 2 in [1].

On the other hand, the discrepancy between the HR-TEM measurement of the Nb core estimated via TEM imaging ($w_{Nbexp}$ =15 nm as in figure 1d in [1], where $w_{Nbexp}$ is equivalent to $2d_S$ in the foregoing) and the values returned by the fit of the data with the QPS model ($w_{NbQPS} \sim 2$ nm as in Table 1 in [1])) can be explained when considering that even if the average thickness of our sample A can be extrapolated as 15 nm, small fluctuations in the Nb thickness or in the Ga penetration along the wire length (100 nm) can be reasonably expected. Such bottlenecks within the nanowire will be dominating the detected QPS and consequently will be sampled by a fit based on a QPS model instead of average properties. It can also be noted that the superconducting order parameter in a proximity system is reduced across the boundary between the two materials of said system and thus a model that does not explicitly account for the proximity

(like the one we used) will be likely to see the nanowire as a single SC with an effective reduced thickness. Here we like to stress that we have also 5-7 nm of error in the $w_{Nbexp}$ obtained from TEM data as show in Fig. 1d. Therefore the calculated $w_{NbQPS}$ ~ 2 nm for device A is in reasonable agreement with the thinnest value $w_{Nbexp}$= 15-7= 8 nm obtained from TEM images, which is likely to dominate the superconducting properties as discussed heretofore. Moreover said agreement appears to be in line or even stronger with respect to the one obtained in similar fits by others groups and discussed in ref. 4 and in ref. 5 of our paper, especially when considering that differences of up to a few orders of magnitudes were found in some of these previous works (see [5] for example).

A. Engel also made further comments [2] on other aspects of our paper, which we will address in the remaining of our reply. The fit of the samples B-D with a LAMH model returned unphysical values for the $T_C$ of Nb as noted by Engels. On the contrary for devices A-D in figure 1 and in Table 1 of [1], there is quite a good agreement between the best-fit values for the normal state resistance and the experimental ones (see for example: $R_{Nexp}(A) \approx 15$ Ω ~ $R_{NFIT}(A) \approx (13+/-1)$ Ω or $R_{Nexp}(C) \approx 2.2$ Ω ~ $R_{NFIT}(C) \approx (2.5+/-0.1)$ Ω). The main reason for the anomalous values of $T_C$ returned by the fit of samples B, C and D is that the phase slip model looses accuracy in the description of the behavior of such devices with length >200 nm [6]. On the other hand at present there are no alternatives or improvements to the QPS models we used. In response to the comments of A. Engel concerning the drop in resistance between 6.5 and 6 K that occurs in the longer (L≥ 500 nm) and wider devices, like B, C and D of [1], we believe that such drop is probably to be attributed to the fact that in these wires there is a competition between the phase slip and non-1D superconductivity, where the latter takes the lead at lower $T$. On general grounds, these longer wires are not the main focus of our work and no further consideration was given in our paper.

To further reassure about the occurrence of QPS in our sample, we add measured I-V curves of nanowires with similar Nb core sizes as our samples B-D and a length of 200 nm as illustrated in figure 1. Said I-V curves clearly show steps, which are the fingerprint of phase slip centres in nanowires [5]. As we do not observe steps in samples with larger Nb cores, we can conclude that the superconductivity in our samples can be associated with phase slip phenomenology and that the comment of Engel [2], can yes call for a characterization of the proximity in a Nb/Nb-Ga system which we can set as a next step of our work, but cannot represent a valid alternative explanation of our data.

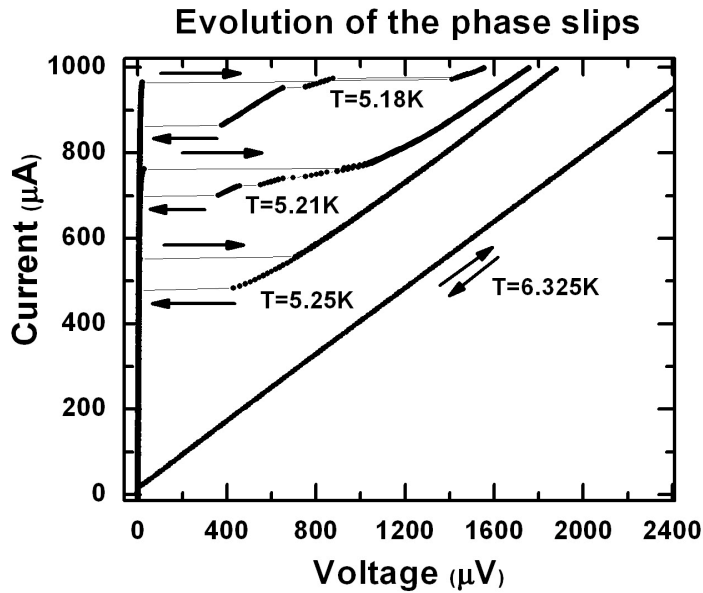

**Figure 1. Current-Voltage characteristics of one of the 200 nm length device. The voltage steps represent the nucleation of successive phase slip centers (see [5-6]).**

# References

[1] G. C. Tettamanzi *et al*, Nanotechnology, **20** (2009) 465302.
[2] A. Engel, submitted to Nanotechnology.
[3] A. Potenza *et al,* Phys. Rev. B, **76 (2007)** 014534.
[4] C. Camerlingo *et al,* Phys. Rev. B, **31 (1985)** 3121, P. D. Scholten and W. G. Moulton, Phys. Rev. B, **15 (1977)** 1318
[5] P. Mikheenko *et al*, Phys. Rev. B, 72 (2005) 174506.
[6] A. Bezryadin, J. Phys.: Condens. Matter, 20 (2008) 043202.